\documentstyle[aps,epsf,twocolumn]{revtex}
\newcommand{\dr}{{\rm d}}
\newcommand{\bea}{\begin{eqnarray}}
\newcommand{\beq}{\begin{equation}}
\newcommand{\eea}{\end{eqnarray}}
\newcommand{\eeq}{\end{equation}}
\begin{document}
\title
{ Exceptional Points of Non-Hermitian Operators}

\author
{W.D. Heiss}

\address
{Institute of Theoretical Physics and Department of Physics,
University of Stellenbosch, 7602 Matieland, South Africa }

\maketitle

\begin{abstract}
Exceptional points associated with non-hermitian operators, i.e.~operators being
non-hermitian for real parameter values, are investigated. The specific characteristics
of the eigenfunctions at the exceptional point are worked out. Within the domain of real 
parameters the exceptional points are the points where eigenvalues switch from 
real to complex values. These and other results are exemplified by a 
classical problem leading to exceptional points of a non-hermitian matrix.
\end{abstract}
\vspace{0.1in}

\vspace{0.1in}

\section{Introduction}
Exceptional points are branch point singularities of the spectrum and 
eigenfunctions, which occur generically when a matrix,
or for instance a Hamilton operator, is analytically continued in a
parameter on which it depends. The term `Exceptional Points' 
has been introduced by Kato \cite{kato}.
When a physical problem is formulated by $H_0+\lambda H_1$ with $\lambda $ being
a strength parameter, the spectrum and eigenfunctions -- $E_n(\lambda )$ and
$|\psi _n(\lambda )\rangle$ -- are in general analytic functions of $\lambda $.
At certain points in the complex $\lambda $-plane two energy levels coalesce. Such
coalescence is not to be confused with a genuine degeneracy, since the eigenspace
of the two coalescing levels is not two but only one dimensional; in fact the 
corresponding eigenvectors also coalesce and there is no two dimensional subspace 
as is the case for a genuine degeneracy.

If both operators, $H_0$ and $H_1$, are hermitian, these singularities -- the 
exceptional points (EP) -- can occur only for complex parameter values $\lambda $. 
As a consequence, at an EP the 
full problem $H_0+\lambda H_1$ is no longer hermitian as such, but when dealing with
matrices, it is still complex symmetric. These cases have been 
studied in some detail
\cite{bendb,hesa} and here we quote the major results.

EPs are always found in the vicinity of a level repulsion.
Suppose that two levels show avoided level crossing when $\lambda$ is 
varied along the real axis; then the analytic continuation into the 
complex $\lambda $-plane yields a complex conjugate pair of EPs where
the two coalescing levels are analytically connected by a square root
branch point \cite{he99}. The 
occurrence of EPs is not restricted to repulsions of bound states, 
a recent paper deals with the repulsion of resonant states \cite{mond}.
Being singularities in the interaction strength EPs determine
the convergence radius of approximation schemes in the theory of effective
interactions \cite{schu}. Quantum mechanical phase transitions are characterised
by a multitude or accumulation points of EPs \cite{hemu,rich}.

There are a number of phenomena, where the physical effect of an EP 
has been at least indirectly observed. Laser induced ionisation states  
of atoms \cite{lat} are a clear manifestation of an 
EP even though in \cite{lat}, it has not been analysed as such. 
A recent theoretical paper \cite{shuv} 
shows that, for a suitable choice of parameters associated with an EP, 
the only acoustic modes in an absorptive medium are circular polarized 
waves with one specific orientation for a given EP. Similarly in
optics, experimental observations in absorptive media \cite{panch} 
reveal the existence of handedness
since the stable mode of light propagation is 
either a left or a right circular polarized wave for appropriately chosen 
parameters. This has been interpreted in \cite{ber2} in  
terms of EPs. A particular resonant behavior of atom waves in crystals  
of light \cite{ober} has been interpreted \cite{ber3}
in terms of EPs. While absorption is essential in all cases, 
some situations clearly point to a chiral behavior of an EP. In fact,
the wave function at the EP has been shown to have definitive
chiral character \cite{heha} and this has been experimentally confirmed
recently \cite{dembrec}.
  
In a previous experiment \cite{demb}, EPs have been investigated in a flat 
microwave cavity. Major findings have been the confirmation of a forth order
branch point of the coalescing wave functions and -- depending on the path in the
complex $\lambda $-plane -- level avoidance associated with width crossing or
level crossing with width avoidance. 
These results are the consequence of the topological structure 
of Riemann sheets at a branch point \cite{he99}. The experiment thus showed that 
this topology is a physical reality.  

In the following we carry the analysis further in that we investigate the
EPs of $H_0+\lambda H_1$ when $H_0$ or $H_1$ or both are no 
longer hermitian. As a consequence, even for real values of $\lambda $, the
problem $H_0+\lambda H_1$ is no longer hermitian. This lack of hermiticity
is different in nature from that discussed above where
dissipation -- like for instance in the optical model in nuclear
physics \cite{mw} -- makes 
$H_0+\lambda H_1$ non-hermitian for complex $\lambda $. 
There is a great variety of problems in the 
literature where either the perturbation or the full Hamiltonian is 
non-hermitian. Boson mapping \cite{sch}, effective interactions \cite{bar}
and the the random phase approximation (RPA)
in many body theory \cite{rs} yield non-hermitian operators. 
More recently a wider class of non-hermitian
Hamiltonians has been proposed to address specific symmetries \cite{ben80} 
or transitional points in specific delocalisation models \cite{hn}. 
These suggestions have led to
a further thorough study \cite{ks} of non-hermitian operators.

The present study is motivated by the
classical problem of two coupled damped oscillators. It gives rise to
non-hermitian matrices in a natural way. The problem is stated in the following
section. The ensuing general treatment of section three yields new 
insights and special
features regarding level repulsion. It is shown that the change from a
complex to a pure real spectrum of a (real) non-hermitian matrix under
variation of the (real) parameter $\lambda $ is due
to the occurrence of a real EP.  As expected, the coalescence at the EP
of two complex eigenvalues into one real eigenvalue 
(which then bifurcates in two real eigenvalues)
yields only one eigenfunction in contrast to the usual two for a genuine
degeneracy. A typical example is the
instability point of the RPA.
In addition, the pattern of level repulsion is distinctly 
different from that of a hermitian problem: the levels approach each 
other in the form of a cusp and not in a smooth way as is the case for
hermitian $H_0$ and $H_1$. These general findings are illustrated in section
four where the example of section two is resumed.

We stress that the present paper focusses upon EPs and not
on the study of non-hermitian operators as such. A summary and
suggestion is given in the last section.

\section{Two coupled damped oscillators}

As a first illustration we consider a simple classical case of two damped
coupled oscillators in one dimension. Denoting by $p_1,p_2,q_1,q_2$ the momenta
and spatial coordinates of two point particles of equal mass the equations of
motion read for the driven system
\beq
{\dr \over \dr t}\pmatrix{p_1 \cr p_2 \cr q_1 \cr q_2}={\cal M}
\pmatrix{p_1 \cr p_2 \cr q_1 \cr q_2} 
 + \pmatrix{c_1 \cr c_2 \cr 0 \cr 0}\exp (i\omega t)
\label{eom}
\eeq
with 
\beq
{\cal M}=\pmatrix{-2g -2k_1 & 2g & -f-\omega_1^2 & f \cr
2g & -2g -2k_2 & f & -f-\omega_2^2 \cr
1 & 0 & 0 & 0 \cr
0 & 1 & 0 & 0 } 
\label{mat}
\eeq
where $\omega _j - ik_j,\; j=1,2$ are essentially the damped frequencies without coupling 
and $f$ and $g$ are the coupling spring constant and damping of the coupling,
respectively. The driving force is assumed to be oscillatory with one single
frequency and acting on each particle with amplitude $c_j$. 
Here we are interested only in the stationary solution being the
solution of the inhomogeneous equation which reads
\beq
\pmatrix{p_1 \cr p_2 \cr q_1 \cr q_2}=(i\omega -{\cal M})^{-1}
\pmatrix{c_1 \cr c_2 \cr 0 \cr 0}\exp (i\omega t).
\label{sol}
\eeq
Resonances occur for the real values $\omega $ of the complex solutions of the
secular equation
\beq
\det |i\omega -{\cal M}|=0
\label{det}
\eeq
and EPs occur for the complex values $\omega $ where
\beq
{\dr \over \dr \omega}\det |i\omega -{\cal M}|=0
\label{der}
\eeq
is fulfilled simultaneously together with Eq.(\ref{det}). We choose the
parameter $f$ as the second variable needed to enforce the simultaneous solution
of Eqs.(\ref{det}) and (\ref{der}) and keep the other parameters of ${\cal M}$ fixed,
but of course any other preference -- like choosing $g$ -- would be just as good
and not alter the essential results. Thus we encounter the problem of finding
the EPs of the matrix problem
\beq
{\cal M}_0+f {\cal M}_1  \label{fep}
\eeq
with
\beq
{\cal M}_0=\pmatrix{-2g -2k_1 & 2g & -\omega_1^2 & 0 \cr
2g & -2g -2k_2 & 0 & -\omega_2^2 \cr
1 & 0 & 0 & 0 \cr
0 & 1 & 0 & 0 } 
\eeq
and
\beq 
{\cal M}_1=\pmatrix{0&0&-1&1\cr 0&0&1&-1
\cr 0&0&0&0 \cr 0&0&0&0 }.
\eeq
Note that ${\cal M}_0$ and ${\cal M}_1$ are not symmetric.
Before we turn to explicit solutions and characteristics of Eqs.(\ref{det}) and 
(\ref{der}) we first address the general problem of EPs of
non-hermitian matrices.

\section{General non-hermitian case}
Like for the hermitian operators the behavior around an EP can be described
locally by a $2\times 2$ matrix. The reduction of an $N$ dimensional
to a two dimensional problem is given below. We always consider a 
situation where the unperturbed problem, denoted by $H_0$, or $h_0$ for
the two dimensional case, is assumed to be diagonal. 
At first we discuss the two dimensional case and assume that the
Jordan decomposition of the non-hermitian perturbation $h_1$, i.e.~$h_1=SJS^{-1}$,
yields a diagonal matrix $J$. We thus consider
\beq
h(\lambda )=\pmatrix{\epsilon _1 &0 \cr 0 & \epsilon _2}+
\lambda S\pmatrix{\omega_1 &0\cr 0& \omega _2}S^{-1}
\label{two} \eeq
with
\beq  S=\pmatrix{\cos \phi _1 & -\sin \phi _2 \cr \sin \phi _1 & \cos \phi _2}. \label{sj} \eeq
Note that for $h_1$ to be symmetric we would have $\phi _1=\phi _2$, i.e.~$S$ would be orthogonal. 
For convenience we have exploited the freedom to use normalized column vectors in $S$.
The two eigenvalues of $h$ are given by
\beq
E_{1,2}(\lambda )={1\over 2}(\epsilon _1+\epsilon _2+\lambda (\omega _1+\omega _2)\pm D)
\label{eiv} \eeq
with the discriminant
\bea D=\bigg((\epsilon _1-\epsilon _2)^2+\lambda ^2(\omega _1-\omega _2)^2+ \qquad \qquad & \cr
2 \lambda (\epsilon _1-\epsilon _2)(\omega _1-\omega _2)\cos (\phi _1+
\phi _2)\sec (\phi _1-\phi _2)\bigg )^{1/2}.&
\label{discr} \eea
The two levels coalesce when $D=0$ that is for
\bea
\lambda _{EP}^{\pm }=-{\epsilon _1-\epsilon _2 \over \omega _1-\omega _2}
\bigg(\cos (\phi _1+\phi _2) & \cr
\pm i \sqrt{\sin 2\phi _1\sin 2 \phi _2}\bigg)&\sec (\phi _1-\phi _2).
\label{ep} \eea

\begin{figure}
\epsfxsize=3.2in
\centerline{
\epsffile{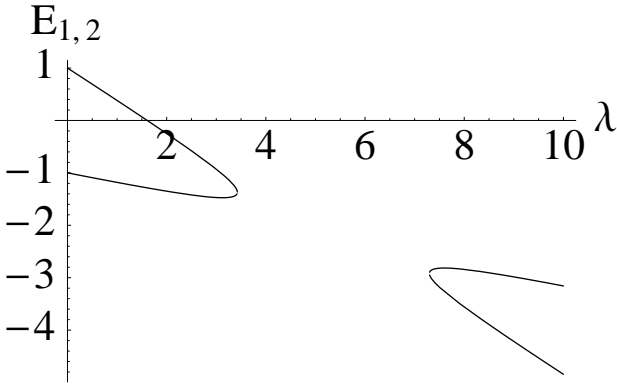}}
\vglue 0.15cm
\caption{Real spectrum in arbitrary units as a function of $\lambda $. In terms of Eq.(\ref{two})
the parameters are $\epsilon _1=-1,\epsilon _2=1,\omega _1=-0.2,\omega _2=-0.6,
\phi_1=-2^0,\phi _2=45^0$. The spectrum is complex between the
EPs at $\lambda_{EP}^+ =7.3\ldots$ and $\lambda _{EP}^-=3.4\ldots$.}
\label{fig1}
\end{figure}

Note that even when all parameters are real the two EPs can now occur on
the real axis. It happens when the signs of $\phi _1$ and $\phi _2$ are different.
The implication is that the spectrum is no longer real when $\lambda $ lies between 
$\lambda _{EP}^-$ and $\lambda _{EP}^+$. In Fig.1 we display a typical case of a
spectrum of that nature 
\footnote{in Fig.1 of \cite{ben80} the spectra $E_k(N)$ exhibit manifestations
of real EPs in the variable $N$.}. We recall that at the EPs, where the
real spectrum ends or begins, only {\sl one} eigenfunction exists of the two by two
matrix problem; its precise form is given below. Here we stress a general property
of a matrix at an EP:
the Jordan decomposition of $h(\lambda _{EP})=TJT^{-1}$ yields a non-diagonal matrix $J$
given by the standard form
\begin{equation}
J_{{\rm EP}}=\pmatrix{ E(\lambda _{EP}) & 1 \cr 0 & E(\lambda _{EP}) }
\label{jordep}
\end{equation}
whereas for all points $\lambda \ne \lambda _{EP}$ the matrix $J_{\lambda }$ is 
diagonal and reads
\begin{eqnarray}
J_{\lambda }&=& T^{-1}h(\lambda )T  \nonumber \\
&=&\pmatrix{ E_1(\lambda ) & 0 \cr 0 & E_2(\lambda ) }.
\label{jord}
\end{eqnarray}
We recall that $T$ is orthogonal (or unitary) only if $h(\lambda )$ is hermitian.

While it is well known that a non-hermitian operator can have a non-real spectrum, 
the deviation from the hermitian case has, for the real part of the spectrum, 
distinct consequences for the shape of level
repulsions; this is exemplified in the following section where typical results of
the two oscillators introduced in the previous section are presented.

Yet the local behavior at the EP is basically the same as for the hermitian case. 
It is clear from Eqs.(\ref{eiv})
and (\ref{discr}) that the two eigenvalues are connected at the square root branch points
situated at $\lambda =\lambda _{EP}^{\pm }$ just as in the hermitian case. The difference
arises in the eigenfunction at the EP. Recall that the coalescence of two eigenvalues at the
EP is not to be confused with a true degeneracy in that there is only one eigenfunction 
at the EP. At $\lambda _{EP}^{\pm }$ this single and unique 
eigenfunction (up to a possible common factor) turns now out to be 
\beq
|\psi _{EP}^{\pm}\rangle =\pm i\sqrt{{\cot \phi _1\over \cot \phi _2}}|1\rangle +
|2\rangle .
\label{eigfr}
\eeq
We note that, in contrast to the symmetric case, the
left hand eigenfunction at the EP (or the eigenfunction of the adjoint problem) is 
now different and reads
\beq
\langle \psi _{EP}^{\pm}|=
\pm i\sqrt{{\tan \phi _1\over \tan \phi _2}}\langle 1| +\langle 2|. \label{eigfl}
\eeq
As a result, the relation
\beq
\langle \psi _{EP}^+|\psi _{EP}^+\rangle =
\langle \psi _{EP}^-|\psi _{EP}^-\rangle =0 \label{scpr}
\eeq
still prevails just as in the symmetric case. The basis vectors $|j\rangle ,\,j=1,2$
refer to the eigenstates of $h_0$.

We only mention the special case where either $\phi _1$ or $\phi _2$ assume the value $0$
or $\pi /2$: in contrast to the hermitian case the confluence of the two EPs does not
invoke a true degeneracy with two independent eigenvectors even though it gives rise -- for
real parameters -- to a real level crossing; also Eq.(\ref{scpr}) is upheld at such points.

To summarize:
for real matrix elements and different signs of the 
angles, the spectrum is complex between the two real EPs. 
Regarding the wave function the quotient of the amplitudes of the two 
coalescing wave functions deviates from that of the hermitian case. 
The genuinely complex superposition of the eigenfunction 
at the EP remains, however. For real angles $\phi _j$ (of equal sign), the fixed 
phase difference of $\pm \pi /2$ between the basis states at $\lambda _{EP}^{\pm }$ 
occurs just as in the symmetric case, but the ratio of the modulus of the
amplitudes is in general not equal to unity. In addition, as the angles 
may be complex, not only the ratio of the modulus
but also the phase difference can be different from the hermitian case. The important
point is, however, that Eq.(\ref{eigfr}) describes, up to a common factor, the only
possible eigenmode at the EP. 

The reduction locally of an $N$-dimensional problem to the appropriate
effective two-dimensional problem
around an EP is, {\it mutatis mutandis}, achieved along the same lines as for hermitian 
operators \cite{heha}. Owing to their vanishing norm
the two coalescing eigenfunctions dominate the complete set of all $N$
normalized eigenfunctions in the immediate vicinity of an EP. 
The expansion of the $N$-dimensional vector 
\begin{equation}
|\psi _{EP}^{\pm }\rangle = 
\sum \beta _k^{\pm }(\lambda )|\chi _k(\lambda )\rangle ,
\label{twod}
\end{equation}
in terms of the complete bi-orthogonal set
$$ \sum |\chi _k(\lambda ) \rangle \langle \chi _k(\lambda )|=1 
\qquad \lambda \ne \lambda _{EP} $$
with $\beta _k^{\pm }(\lambda )=\langle \chi _k(\lambda )|
\psi _{EP}^{\pm }\rangle $, 
contains virtually only two terms for $\lambda \to \lambda _{EP}$.
In fact we may write
\beq
\lim _{\lambda \to \lambda _{EP}} \pmatrix{\beta _1^{\pm } \cr \vdots \cr 
\beta _N^{\pm }}=
\pmatrix{0 \cr \vdots \cr \pm i\sqrt{{\cot \phi _1\over \cot \phi _2}} \cr 
1 \cr 0 \cr \vdots }
\label{restr}
\eeq
up to a common factor; the two non-zero positions are given by the values 
$k,k+1$ for which
$|\chi _k(\lambda )\rangle $ and $|\chi _{k+1}(\lambda )\rangle $ coalesce. From
Eqs.(\ref{twod},\ref{restr}) the effective two dimensions for any 
$|\psi (\lambda )\rangle $ becomes obvious within a small neighbourhood
of $\lambda _{EP}$.

We do not discuss cases where the Jordan decompositions of $h_0$ or $h_1$ or both
do not yield diagonal but block matrices as this does not affect the local
behavior at an EP. This should not be confused with the fact 
that in all cases an EP 
of the full problem $H_0+\lambda H_1$ (or $h_0+\lambda h_1$) is characterized by a 
non-diagonal matrix $J$ (see Eq.(\ref{jordep})) of its Jordan decomposition. 

\section{Examples}
While there are various physical reasons to consider non-hermitian operators, we here focus
on the simple mechanical model introduced in section two. Note that the model can be
easily translated into a corresponding electronic setting using two coupled R-L-C circuits. EPs can always be found for some complex
values of the pair $(\omega ,f)$, but a complex value of the spring constant $f$ does not appear
physical. This is in contrast to quantum mechanical cases discussed previously \cite{he99} where
dissipation is often described by an effective complex interaction. In the classical
model we therefore introduce the damping term of the coupling denoting its strength by 
the real constant $g$. For given values of $\omega _j$ and $k_j$ we determine $g$ such that
an EP occurs at a real value of $f$. The associated two coalescing energies
are then complex describing a damped oscillation being sustained by the driving
force.

\begin{figure}
\epsfxsize=3.2in
\centerline{
\epsffile{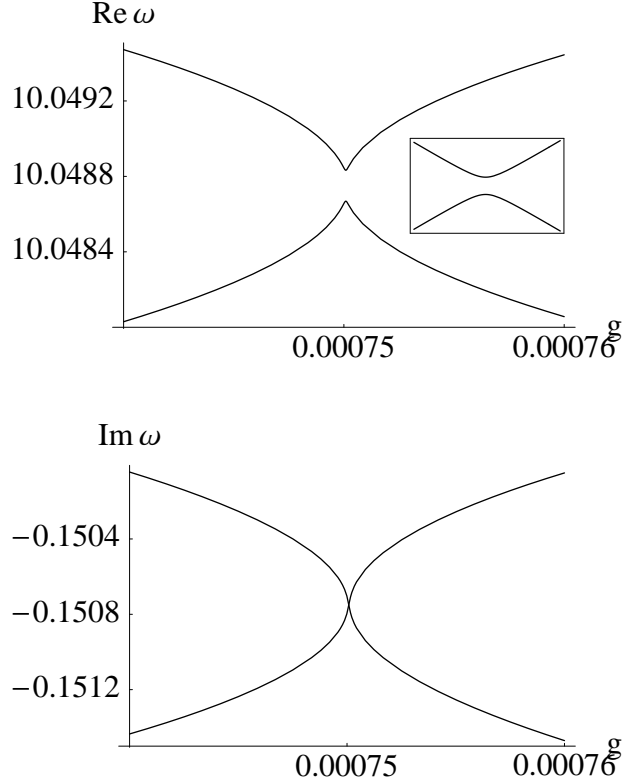}}
\vglue 0.15cm
\caption{ Real and imaginary parts of two repelling levels as a function of the
coupling damping $g$. The parameters are $\omega _1=\omega _2=10,\;k_1=0.2,
\;k_2=0.1$. The spring constant $f=1.005$ is chosen such that an EP occurs at 
$g=0.00075$ close to the real axis. The units are arbitrary. The inset illustrates
schematically a typical level repulsion for a symmetric matrix}
\label{fig2}
\end{figure}

The particular model reduces the resultant of Eqs.(\ref{det}) and (\ref{der}) to a 
polynomial of fifth order in $f$ which is readily solved. The symmetry 
of the model implies that an EP at the pair $(\omega _{EP},f_{EP})$ is always 
associated with an EP at $(-\omega ^*_{EP},-f_{EP})$; we focus our 
attention on positive $f$, i.e.~a repulsive spring, and the physical
requirement $\Im \omega _{EP}<0$ implying proper damping. Obviously,
EPs can occur only if either $\omega _1\ne \omega _2$ or $k_1\ne k_2$ or both as
otherwise a genuine degeneracy is found for $f=g=0$.

To get a good understanding for the EPs we first turn our attention to the behavior
of the eigenvalues of ${i\cal M}$ in Eq.(\ref{mat}) as functions of $f$ and $g$;
the eigenvalues are the solutions of Eq.(\ref{det}).
In Fig.2 the real and imaginary parts of two levels coalescing at an EP very close to
the real $g$-axis are plotted {\it versus} $g$ using for $f$ a real fixed value 
chosen such that an EP occurs in the vicinity of $g=0.00075$.

\begin{figure}
\epsfxsize=3.2in
\centerline{
\epsffile{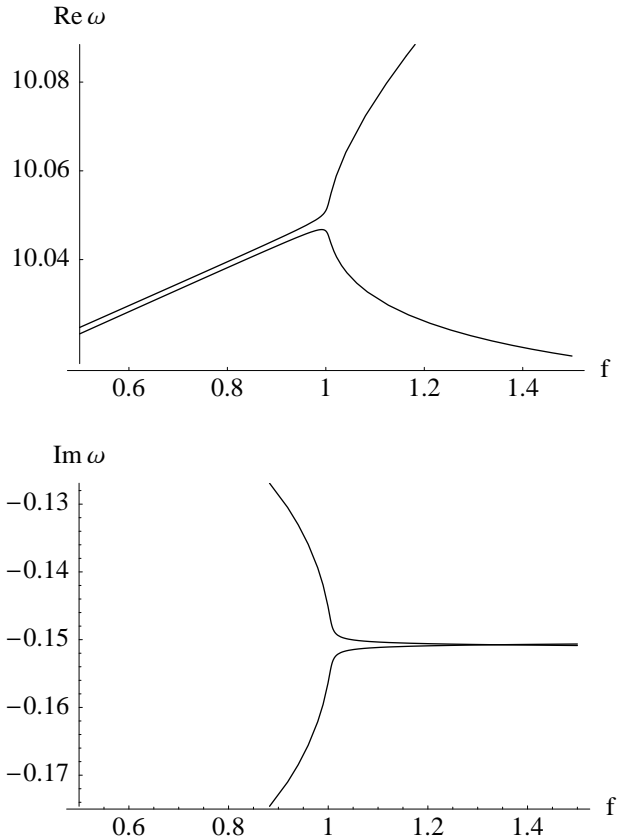}}
\vglue 0.15cm
\caption{ Real and imaginary parts of two repelling levels as a function of the
spring constant $f$ for fixed value $g=0.00075$. The parameters are the same as 
in Fig.2}
\label{fig3}
\end{figure}

The level repulsion of the real parts and the expected crossing 
\cite{he99} of the imaginary parts are distinctly different in shape 
from the usual appearance for hermitian matrices. The cusp
originates from the plain square root behavior of the singularity, i.e.~the 
difference between the two levels is controlled by 
$\sim \sqrt{\lambda -\lambda _{EP} }$ 
(in Fig.2 $\lambda \equiv g$); this is in contrast to the two complex 
conjugate EPs occurring in the hermitian case where this difference
is controlled by 
$\sim \sqrt{(\lambda -\lambda _{EP})(\lambda -\lambda ^* _{EP})}=
\sqrt{(\lambda -\Re \lambda _{EP})^2+(\Im \lambda _{EP})^2}$ and hence 
produces a smooth approach. 
For illustration a typical shape of the latter is drawn schematically in the inset. 
The deviation from a hermitian case is even more dramatic when the
two levels are plotted against $f$ for a fixed $g=0.00075$ as illustrated in 
Fig.3. Yet, the pattern is understood by the same mechanism being a square root branch cut 
running along the real $f$-axis and having a branch point at $f\approx 1$.

As we deal with a classical system we now turn to the behavior of the complex
amplitudes $q_1(\omega )$ and $q_2(\omega )$ of Eq.(\ref{sol}). The overall 
oscillatory time behavior is of no interest, we rather concentrate on the modulus 
and the phase difference. In general these complex amplitudes depend on the amplitudes $c_j$
of the driving force by Eq.(\ref{sol}). However, as discussed in the previous section, at the
EP there is only one mode possible given by Eq.(\ref{eigfr}) up to a global constant factor.
In other words, at the EP the ratio of the amplitudes of the two coalescing modes is given by 
$i\sqrt{\cot \phi _1/\cot \phi _2}$ which is a function of only 
${\cal M}_0$ and ${\cal M}_1$ and is independent of a driving force,
i.e.~of the $c_j$. At close distance to 
$\omega _{EP}$ the correct value for the ratio must therefore approximately be attained.

\begin{figure}
\epsfxsize=3.2in
\centerline{
\epsffile{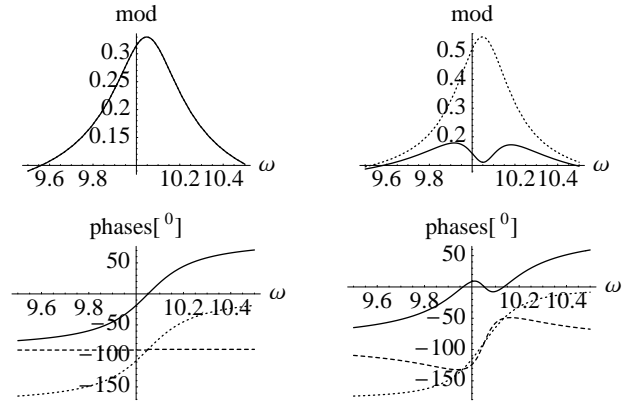}}
\vglue 0.15cm
\caption{Top: Modulus of the amplitudes $q_1(\omega )$ (solid line) and 
$q_2(\omega )$ (dotted line) for different driving amplitudes (see text); 
$|q_1(\omega )|$ agrees perfectly with $|q_2(\omega )|$ on the top left.
Bottom: the respective phases indicated in degrees. The dashed line is the 
phase difference between the amplitudes, for convenience the negative value
is plotted. The parameters are the same as in 
Fig.2}
\label{fig4}
\end{figure}

This is well demonstrated in Fig.4 where the moduli and phases are plotted against the
driving frequency for the same parameters as in the previous figures giving rise to an 
EP at $\omega _{EP} =10.05-0.15\,i$. The top drawings are the moduli of the amplitudes
with the left one referring to $c_1=i,\,c_2=1$ and the right one to $c_1=-i,\,c_2=1$;
below are the respective phases. The former choice (left column in Fig.4) is driving the two masses
with equal strength but with a leading phase of $\pi /2$ for the first mass. From
the drawing we see that $q_1(\omega )/q_2(\omega )\approx +i$ through the whole resonance.
This value is almost equal to the exact value being 
$q_1(\omega _{EP})/q_2(\omega _{EP})=0.0049+i\, 1.000\ldots$ 
(see (\ref{exact}) below). 
If, however, the `incorrect' input is enforced like the lagging
phase (right column), there is more variation in the response. Yet, at the resonance
the phase difference still is $+\pi /2$, i.e~opposite to the driving force and in line
with the mode at the EP, even though the ratio of the moduli is quite different from unity.
If the driving force is getting closer to $\omega _{EP}$, i.e.~if a slightly damped
excitation is used, the ratio does approach the exact value {\sl irrespective} of
the values $c_j$. In fact, after some slightly tedious but straightforward algebra the result
\bea
{q_1(\omega _{EP})\over q_2(\omega _{EP})}={p_1(\omega _{EP})\over p_2(\omega _{EP})}= 
\qquad \qquad \qquad  \qquad \cr
={f_{EP}+\omega _2^2-2i(g+k_2)\omega _{EP}-\omega _{EP}^2\over f_{EP}-2ig\omega _{EP}}
\qquad \cr \qquad
={f_{EP}-2ig\omega _{EP}\over f_{EP}+\omega _1^2-2i(g+k_1)\omega _{EP}-\omega _{EP}^2 }
\label{exact}
\eea
is obtained yielding the numerical value indicated above for the 
parameters considered.
While this result is obtainable analytically for the particular case of Eq.(\ref{fep}),
in general one has to resort to the two dimensional reduction by numerical means
and then use Eqs.(\ref{two}), (\ref{sj}) and (\ref{eigfr}) to find the
amplitude ratio. 

\section{Summary}
The physical relevance of EPs and their observability has been
presented in the introduction. Consideration of a simple mechanical problem
leads to non-hermitian matrices and the study of the associated EPs
has produced new general insights. The parameters for which a real spectrum switches
to complex values is clearly related to the occurrence of EPs on the
real axis. The instability point of the RPA is just one case in point. 
In addition, the specific shape of the spectrum
can be quite different from the one encountered for symmetric matrices. On the
other hand, the topological structure, i.e.~the Riemann sheet structure of the
energy surfaces is independent of whether $H_0$ and/or $H_1$ are hermitian or not. 
The eigenfunctions at the EP have a structure
similar to the symmetric case except for the value of the ratio of the two
relevant states. This changes from $\pm i$ for the symmetric case to 
$\pm i\sqrt{\cot \phi _1/\cot \phi _2}$ for the non-hermitian case. Note also, that
this ratio is different for the left hand eigenfunction at the EP.

The universal significance of the EPs is once more underlined by 
the particular example from classical physics.
While the general features of EPs for non-hermitian $H_0$ and $H_1$ 
have been presented in section three, we believe that the particular results 
of section four can be experimentally confirmed, results whose 
analogues have so far been implemented in sophisticated 
microwave cavities \cite{dembrec,demb}, optical systems \cite{panch} and atomic 
spectra \cite{lat}.

\vskip 1cm

{\bf Acknowledgment} The author acknowledges critical comments from Hendrik Geyer
at the Department of Physics of the University of Stellenbosch.


\begin{thebibliography}{99}
\bibitem{kato}T.~Kato, \emph{Perturbation theory of linear operators}
(Springer, Berlin, 1966).

\bibitem{bendb} C.M.~Bender and S.A.~Orszag {\it Advanced Mathematical  
Methods for Scientists and Engineers}, McGraw-Hill 1978 
 
\bibitem{hesa}W.D.~Heiss and A.L.~Sannino, J.~Phys.~A {\bf 23}, 1167 (1990);
Phys.~Rev.~A {\bf 43}, 4159 (1991); W.D.~Heiss,
Phys.~Rep. {\bf 242}, 443 (1994).

\bibitem{he99}W.D.~Heiss, Eur. Phys. J. {\bf D 7}, 1 (1999);
Phys. Rev.~E {\bf 61}, 929 (2000).

\bibitem{mond} E.~Hernandez, A.~Jaureguit and A.~Mondragon,  
J. Phys.~A {\bf 33}, 4507 (2000) 

\bibitem{schu} T.H.~Schucan and H.A.~Weidenm\"uller, Ann.Phys.(NY) {\bf 73}, 108 (1972)

\bibitem{hemu} W.D.~Heiss and M.~M\"uller
Phys.~Rev.~{\bf E66}, 016217 (2002)

\bibitem{rich} J. Richert, arXiv:quant-ph/0209119

\bibitem{lat}O. Latinne \emph{et al.},
  Phys.~Rev.~Lett.~{\bf 74}, 46 (1995).

\bibitem{shuv} A.L.~Shuvalov and N.H.~Scott, Acta Mech. {\bf 140}, 1
(2000).

\bibitem{panch} S~Pancharatnam, Proc.Ind.Acad.Sci. {\bf XLII}, 86 (1955) 
 
\bibitem{ber2} M.V.~Berry, Current Science {\bf 67}, 220 (1994) 
 
\bibitem{ober} M.K.~Oberthaler {\it et.al.} Phys.~Rev.~Lett. {\bf 77}, 4980 (1996)  

\bibitem{ber3} M.V.~Berry and D.H.J.~O'Dell, J.Phys.A{\bf 33}, 2093 (1998) 

\bibitem{heha}W.D.~Heiss and H.L.~Harney, Eur. Phys. J. {\bf D 17}, 149 (2001).

\bibitem{dembrec} C~Dembowski {\it et.al.},
Phys. Rev. Lett. {\bf 90}, 034101 (2003)

\bibitem{demb}C.~Dembowski \emph{et al.}, Phys. Rev. Lett. {\bf 86},
787 (2001). 

\bibitem{mw} C.~Mahaux and H.A.~Weidenmuller {\it Shell Model Approach to Nuclear
Reactions}, North-Holland, Amsterdam (1969)

\bibitem{sch} F.G.~Scholtz, H.B.~Geyer and F.J.W.~Hahne, Ann.Phys.(NY) {\bf213}, 74 (1991)

\bibitem{bar} B.B.~Barrett (Ed.) {\it Effective Interactions and Operators in Nuclei},
Lecture Notes in Phyiscs, Vol.40, Springer, Berlin (1975)

\bibitem{rs} P.~Ring and P.~Schuck, {\it The Nuclear Many Body Problem}, Springer, New York (1980)

\bibitem{ben80} C.M.~Bender and S.~Boettcher, Phys.Rev.Lett. {\bf 80}, 5243 (1998)

\bibitem{hn} N.~Hatano and D.R.~Nelson, Phys.Rev.Lett. {\bf 77}, 570 (1996)

\bibitem{ks} R.~Kretchmer and L.~Szymanowski, arXiv:quant-ph/0105054

\end{thebibliography}
\end{document}